\documentstyle[seceq,preprint,epsfig]{jpsj}

\title
{
The rejuvenation effect in the two-state random energy model
}

\author
{Mitsuhiro {\sc Kawasaki}\footnote{E-mail: M.Kawasaki@cmt.phys.kyushu-u.ac.jp}}

\inst{Department of Physics, Kyushu University, Fukuoka 812-8581, Japan}

\recdate
{
\today
}

\abst{
Theoretical analyses of the random energy model with only two states and 
its extension with a hierarchy of only two levels show that 
these models reproduce out-of-equilibrium phenomena observed in 
experiments of glassy materials; the rejuvenation effect (the chaos effect), 
i.e. the abrupt jump and subsequent relaxation 
of the out-of-phase susceptibility 
as if the system rejuvenates when the temperature is lowered, and the 
power-law relaxation of the two-time correlation function. 
Our results suggest that also in an assembly of small systems with 
relaxation times distributed broadly some of these interesting 
out-of-equilibrium phenomena can be observed.
}

\kword
{
random energy model, hierarchical diffusion, rejuvenation, chaos effect, 
reinitialization
}

\begin{document}
\sloppy
\maketitle

\section{Introduction}

The free energy of disordered systems such as spin-glasses, 
structural glasses, polymers and proteins~\cite{Frauenfelder91} 
is considered to have a very complex structure with numerous local minima. 
Due to this complexity, it takes very long time for a system 
to equilibrate and various out-of-equilibrium phenomena are observed 
even in macroscopic time-scales.~\cite{Vincent96}

For example, in the temperature cycling experiment reported for
spin-glasses~\cite{Vincent95}, 
the relaxation of the out-of-phase 
susceptibility $\chi ''$ is measured in the following three stages. 
In the first stage, the sample is quenched from above $T_g$ down to a 
temperature $T$ below $T_g$ and is kept at this temperature for a period
of $t_1$. Then the sample is perturbed by changing the temperature from $T$ to 
$T-\triangle T$ and is kept at $T-\triangle T$ 
for a period of $t_2$ (the second stage). 
After that, the temperature is returned to $T$ in the third stage. 
The effect of the perturbation is examined by comparing the perturbed
data with unperturbed data in the third stage. In the case $\triangle T > 0$, 
both data coincide except at the very beginning of the third stage. 
This is called the memory effect, since the sample remembers the relaxation 
in the first stage. On the other hand, 
an abrupt jump and subsequent relaxation restarted as if the sample 
rejuvenates are observed at the beginning of the second stage, which 
is called the rejuvenation effect or the chaos effect since it 
shows that equilibrium state at $T-\triangle T$ is very different from 
one at $T$, i.e. chaotic dependence of equilibrium state on 
temperature. These memory and 
rejuvenation effect are observed also in a cooling and reheating 
experiment for spin-glasses by Jonason et al.~\cite{Jonason98}

According to the Parisi's mean-field solution of the SK model, 
the free energy of the spin-glass has a very complex structure with 
numerous local minima. From this complex free energy landscape, 
very slow relaxation, aging, the memory effect and the rejuvenation effect 
can be explained by analyzing the 
random energy model and its extension.~\cite{Bouchaud95,Sasaki00}

However, some of these out-of-equilibrium phenomena are features 
not restricted to systems with numerous local minima, 
since it is well-known that the power-law relaxation can be reproduced by 
an assembly of small systems with very different relaxation times. 
Furthermore, when the energies of local minima deviate very strongly from 
valley to valley, equilibrium state, i.e. the Gibbs measure, is 
sensitive to temperature even if the number of local minima is small.
Hence, it is expected that such small systems can show 
the rejuvenation effect due 
to this sensitivity of equilibrium state to temperature.
The aim of this paper is to show that the rejuvenation effect is a
feature not restricted to systems with numerous local minima and 
models with small number of local minima 
can reproduce the slow relaxation and the rejuvenation effect by 
analyzing the random energy model with 
only two states and its extension with hierarchy of only two layers.

The organization of this paper is the following; 
In section 2 we present the models analyzed in this paper. 
In section 3 we show that the two-time correlation function decays in 
power-law. In section 4 we show that the rejuvenation effect is reproduced by 
analyzing the out-of-phase susceptibility. 
Finally our conclusion and discussion about 
the fluctuation-dissipation theorem are presented in section 5.

\section{The models}

In this section, we describe two models analyzed in this paper; 
the two-state random energy model (2S-REM) 
and the two-layer random energy model (2L-REM).

\subsection{The two-state random energy model}

\begin{figure}
\centerline{\epsfxsize=8cm
\epsffile{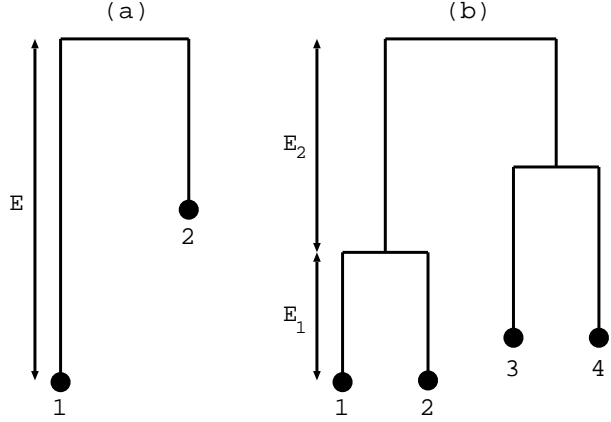}
}
\caption{The structure of the two-state random energy model (a) and the 
structure of the two-layer random energy model (b). 
The closed circles represent accessible states.}
\label{fig1}
\end{figure}

At first we explain the 2S-REM schematically shown in Fig.~\ref{fig1} 
where only two states exists 
and the state energies $E$ are independent random 
variables identically distributed as 
\begin{equation}
\rho (E) = \left\{ 
\begin{array}{ll}
\frac{1}{k_B T_c} \exp(-E/k_B T_c) & \mbox{when $E \geq 0$} \\
0                                  & \mbox{otherwise},
\end{array}
\right.
\label{dist_E}
\end{equation} 
where $T_c$ is a parameter to express the extent of dispersion of $E$.
From the Arrhenius law, the transition rate $w_i$ from site $i$ is related 
to $E$ as 
\begin{equation}
w_i = \frac{1}{\tau_0} \exp(-E/k_B T).
\label{jump_rate} 
\end{equation}
where $\tau_0$ is a microscopic time scale. From eq.~(\ref{dist_E}), it is 
shown that the distribution of the transition rate is given as 
\begin{equation}
P(w) = \left\{ 
\begin{array}{ll}
\alpha w^{\alpha-1} & \mbox{when $0 < w \leq 1$} \\
0                   & \mbox{otherwise},
\end{array}
\right.
\end{equation}
where $\alpha \equiv T/T_c$ and the microscopic time scale $\tau_0$ is set to 
the time unit. 
Referring to the two states as $1$ and $2$, 
the master equation for the probabilities $P_i(t)$ that the system is found 
at the state $i$ is written by the transition rates as 
\begin{equation}
\frac{d}{dt} \left( \begin{array}{c} P_1(t) \\ P_2(t) \end{array} \right) = 
\left( \begin{array}{cc} -w_1 & w_2 \\ w_1 & -w_2 \end{array} \right) 
\left( \begin{array}{c} P_1(t) \\ P_2(t) \end{array} \right).
\label{master1}
\end{equation}
Furthermore, 
we assume that an observable, which we call the ``magnetization'', takes the 
value $-1(+1)$ when the system is in the state $1(2)$.

\subsection{The two-layer random energy model}

The structure of the two-layer random energy model (2L-REM) is 
shown in Fig.~\ref{fig1}. 
We assume that the system is constructed by piling up the 2S-REM two times 
hierarchically. 
Consequently, there are four states, which we refer to as $1, 2, 3$ and $4$, 
where we assume that the state $1(3)$ is directly connected to $2(4)$.
The energy of the n-th layer counted from the bottom, $E_n$, 
is given according to the distribution 
$\rho (E_n) = \exp[-E_n/k_B T_c(n)]/k_B T_c(n)$ and each layer has a different 
$T_c(n)$ chosen so that $T_c(1) < T_c(2)$. 
The master equation for the probabilities that the system is found at 
the state $i$, $P_i(t)$, 
is written by the transition rates as 
\begin{equation}
\frac{d}{dt} \left( \begin{array}{c} P_1(t) \\ P_2(t) \\ P_3(t) \\ P_4(t) 
\end{array} \right) = 
\left( \begin{array}{cccc} 
-w_1-w_1 w_2 & w_1 & w_3 w_4/2 & w_3 w_4/2 \\ 
w_1 & -w_1-w_1 w_2 & w_3 w_4/2 & w_3 w_4/2 \\ 
w_1 w_2/2 & w_1 w_2/2 & -w_3-w_3 w_4 & w_3 \\
w_1 w_2/2 & w_1 w_2/2 & w_3 & -w_3-w_3 w_4 
\end{array} \right) 
\left( \begin{array}{c} P_1(t) \\ P_2(t) \\ P_3(t) \\ P_4(t) 
\end{array} \right),
\label{master2} 
\end{equation}
where $w_1$ and $w_3$ denotes the transition rates from lower layer to 
upper layer and $w_2$ and $w_4$ denotes the transition rates between 
different branches. From the distribution of the state energy $\rho(E_n)$, 
$w_1$ and $w_3$ distributes as $\alpha_1 w^{\alpha_1-1}$ where $\alpha_1 
\equiv T/T_c(1)$ and $w_2$ and $w_4$ distribute as $\alpha_2 w^{\alpha_2-1}$ 
where $\alpha_2 \equiv T/T_c(2)$.
Finally, we assume that the magnetization takes the 
values $-1-\epsilon,-1+\epsilon,1-\epsilon$ and $1+\epsilon$ when the 
system is in states $1, 2, 3$ and $4$.

\section{The long-time behavior of the two-time correlation function}

In this section we analyze the two-time correlation function 
of the ``magnetization'' for the two models. 
We assume that the initial conditions are given as 
\begin{eqnarray}
P_1(0) = P_2(0) = 1/2 \ \mbox{for 2S-REM}, \\ 
P_1(0) = P_2(0) = P_3(0) = P_4(0) = 1/4 \ \mbox{for 2L-REM}, 
\end{eqnarray}
to see the relaxation after quench from infinitely high 
temperature to $T$ below $T_c$.

\subsection{2S-REM}

At first, we calculate the correlation function for the 2S-REM.
The master equation eq.~(\ref{master1}) is solved exactly and the solution is 
given by 
\begin{eqnarray*}
P_1(t) = \frac{(w_1-w_2) \exp[-(w_1+w_2)t]+2 w_2}{2 (w_1+w_2)}, \\ 
P_1(t) = \frac{(w_2-w_1) \exp[-(w_1+w_2)t]+2 w_1}{2 (w_1+w_2)}.
\end{eqnarray*}
The conditional probabilities $G_{ij}(t)$ 
that the system is found at the state $i$ 
at time $t$ if the system is at the state $j$ at time $0$ are given by 
\begin{eqnarray*}
G_{11}(t) = \frac{w_1 \exp[-(w_1+w_2)t]+w_2}{w_1+w_2}, \\
G_{12}(t) = \frac{-w_2 \exp[-(w_1+w_2)t]+w_2}{w_1+w_2}, \\
G_{21}(t) = \frac{-w_1 \exp[-(w_1+w_2)t]+w_1}{w_1+w_2}, \\
G_{22}(t) = \frac{w_2 \exp[-(w_1+w_2)t]+w_1}{w_1+w_2}.
\end{eqnarray*}
With these quantities, the two-time correlation function $C(t,t')$ 
of the magnetization is written as 
\begin{equation}
C(t,t') = \sum_{i,j} m_i m_j <G_{ij}(t-t') P_j(t')>,
\end{equation}
where the angle bracket denotes average over distribution of the jump
rates and $m_i$ denotes the magnetization of the state $i$. 

It is shown that when $t >> 1$(the time unit is $\tau_0$) 
\begin{equation}
\left<\frac{w_1^m w_2^n}{(w_1+w_2)^2} \exp[-(w_1+w_2)t]\right> \simeq 
\frac{\alpha}{2 (2 \alpha+1)}\Gamma(\alpha+m)\Gamma(\alpha+n)t^{-2 \alpha}, 
\label{formula1}
\end{equation}
by using the asymptotic form of the incomplete gamma function 
$\gamma(\alpha+m,t) \simeq \Gamma(\alpha+m)-t^{\alpha+m-1} \exp(-t)$.
Consequently, using eq.~(\ref{formula1}), when $t, t', t-t' >> 1$ the correlation function is written as 
\begin{equation}
C(t,t') \simeq A 
[t^{-2 \alpha}-t'^{-2 \alpha}+2 \alpha (t-t')^{-2 \alpha}]+B, 
\label{c}
\end{equation}
where $A \equiv \alpha^2 \Gamma(\alpha)^2/(2 \alpha+1), 
B \equiv <(w_1-w_2)^2/(w_1+w_2)^2>$.
This expression of the two-time correlation function implies that 
the out-of-equilibrium effect, i.e. the dependence on $t'$, persists after 
very long time and the equilibrium relaxation follows the power-law with 
the exponent $-2 T/T_c$.

\subsection{2L-REM}

Next, we calculate the correlation function for the 2L-REM.
From the symmetry of the probabilities and the conditional probabilities; 
\begin{eqnarray*}
P_1(t)=P_2(t)\equiv P_-(t), P_3(t) = P_4(t) \equiv P_+(t), \\
G_{12}(t)=G_{21}(t)\equiv G_{--}(t), G_{34}(t)=G_{43}(t)\equiv G_{++}(t), \\ 
G_{11}(t)=G_{22}(t)\equiv r_{--}(t), G_{33}(t)=G_{44}(t)\equiv r_{++}(t), \\ 
G_{31}(t)=G_{32}(t)=G_{41}(t)=G_{42}(t)\equiv G_{+-}(t), \\ 
G_{13}(t)=G_{14}(t)=G_{23}(t)=G_{24}(t)\equiv G_{-+}(t),
\end{eqnarray*}
the correlation function is rewritten in terms of the probabilities and 
the conditional probabilities defined above, 
$P_-(t),P_+(t),G_{++}(t),r_{++}(t),G_{+-}(t),G_{-+}(t)$, as 
\begin{eqnarray}
C(t,t') = 4 (1+\epsilon^2) <r_{--}(t-t')P_-(t')> & 
	  +4 (1-\epsilon^2) <G_{--}(t-t')P_-(t')> \nonumber \\
	&  -8<G_{-+}(t-t')P_+(t')>.	  
\end{eqnarray}
By solving the master equation eq.~(\ref{master2}), the following expressions 
for the probabilities and the conditional probabilities are given; 
\begin{eqnarray*}
P_-(t) = \frac{(w_1 w_2-w_3 w_4)\exp[-(w_1 w_2+w_3 w_4)t]+2w_3 w_4}
{4 (w_1 w_2+w_3 w_4)}, \\ 
P_+(t) = \frac{-(w_1 w_2-w_3 w_4)\exp[-(w_1 w_2+w_3 w_4)t]+2w_1 w_2}
{4 (w_1 w_2+w_3 w_4)}, \\ 
r_{--}(t) = \frac{1}{2}\exp[-w_1(2+w_2)t]+
\frac{w_3 w_4+w_1 w_2 \exp[-(w_1 w_2+w_3 w_4)t]}{2 (w_1 w_2+w_3 w_4)}, \\
G_{--}(t) = -\frac{1}{2}\exp[-w_1 (2+w_2)t]+
\frac{w_3 w_4+w_1 w_2 \exp[-(w_1 w_2+w_3 w_4)t]}{2 (w_1 w_2+w_3 w_4)}, \\
G_{-+}(t) = \frac{w_3 w_4-w_3 w_4 \exp[-(w_1 w_2+w_3 w_4)t]}
{2(w_1 w_2+w_3 w_4)}. 
\end{eqnarray*}
By using the following expressions; 
\begin{equation}
<\exp(-w_1 w_2 t)>=\frac{\alpha_1 \alpha_2}{\alpha_1-\alpha_2}
[\Gamma(\alpha_2)-\Gamma(\alpha_2,t)]t^{-\alpha_2}-
\frac{\alpha_1 \alpha_2}{\alpha_1-\alpha_2}
[\Gamma(\alpha_1)-\Gamma(\alpha_1,t)]t^{-\alpha_1}
\end{equation}
and for large $t,t'$ 
\begin{equation}
<\exp(-2w_1 t-w_1 w_2 t')> \simeq \alpha_1 \alpha_2 \Gamma(\alpha_1)
t^{-\alpha_1} \left(\frac{t}{t'}\right)^{-\alpha_2} \int_0^{t'/t}dx 
\frac{x^{\alpha_2-1}}{(2+x)^{\alpha_1}},
\end{equation}
when $t,t',t-t'$ are large the correlation function is consequently given as 
\begin{eqnarray}
C(t,t') = \frac{\alpha_1^2 \alpha_2^3 \Gamma(\alpha_2)^2}
{\alpha_2 (2\alpha_2+1) (\alpha_1-\alpha_2)^2}[t^{-2\alpha_2}-t'^{-2\alpha_2}
& + 2 \alpha_2 (t-t')^{-2\alpha_2}]+2 \epsilon^2 c (t-t')^{-\alpha_1} 
\nonumber \\ 
& +\left<\frac{(w_1 w_2-w_3 w_4)^2}{(w_1 w_2+w_3 w_4)^2}\right>
+O(t^{-\alpha_1-\alpha_2}),
\label{correlation}
\end{eqnarray}
where $c$ is a function of parameters, $\alpha_1$ and $\alpha_2$, defined as 
\begin{eqnarray}
c \equiv \frac{\alpha_1^2 \alpha_2^2 \Gamma(\alpha_1)}{\alpha_2-\alpha_1} 
\int_0^{\infty}ds \int_0^1dx \frac{x^{\alpha_2-1}}{(2+x)^{\alpha_1}} 
& \{\alpha_2 s^{-\alpha_2-1}[\Gamma(\alpha_2)-\Gamma(\alpha_2,s)]
\nonumber \\
& -\alpha_1 s^{-\alpha_1-1}[\Gamma(\alpha_1)-\Gamma(\alpha_1,s)]\}.
\end{eqnarray}

This asymptotic form implies that 
when $2 \alpha_2 < \alpha_1$ ,i.e. $2 T_c(1) < T_c(2)$, 
the dynamics in the upper layer of the hierarchy 
is dominant and the asymptotic form is same as that of 
the 2S-REM except for the difference of constants 
since the leading term of eq.~(\ref{correlation}) is proportional to 
$[t^{-2\alpha_2}-t'^{-2\alpha_2} + 2 \alpha_2 (t-t')^{-2\alpha_2}]$.
On the other hand, when $2 T_c(1) > T_c(2)$ the long-time behavior is 
governed by the dynamics in the lower layer, since the leading term of 
eq.~(\ref{correlation}) is proportional to $(t-t')^{-\alpha_1}$.

\section{The out-of-phase susceptibility and the rejuvenation effect}

In this section we show that the 2S-REM and 2L-REM reproduce the rejuvenation 
effect, which is the abrupt jump and subsequent relaxation 
of the out-of-phase susceptibility when the temperature is lowered.

We explain the definition of the out-of-phase susceptibility. 
Assuming that the magnetization when the system is at the state $i$ is $m_i$ 
and $h(t)$ denotes the applied ``magnetic field'' at time $t$, 
the state energy is changed by $-m_i h(t)$ 
and the transition rate from the state $i$ is modified by the factor 
$\exp[-m_i h(t)/k_B T]$. 
Then the magnetization at time $t$ is given by 
\begin{equation}
m(t) = \sum_i m_i P_i(t).
\end{equation}
We assume that the applied magnetic field oscillates in time as 
\begin{equation}
h(t) = h e^{-i \omega t} \ \mbox{when} \ t \geq 0.
\end{equation}
Then the out-of-phase susceptibility $\chi''$ is defined as the imaginary 
part of the susceptibility 
\begin{equation}
\chi(\omega,t) \equiv \lim_{h \rightarrow 0} \frac{m(t)}{h e^{-i \omega t}}.
\end{equation}

We observe the out-of-phase susceptibility defined in this way in 
the following temperature variation protocol; In the first stage, 
the sample is quenched at time $0$ from infinitely high temperature 
to a temperature $T$ below $T_c$ and 
the oscillating magnetic field is applied at the same time. After that, the 
sample is kept at the temperature $T$ for a period of $t_1$. 
In the second stage, the temperature is changed from $T$ to $T-\triangle T$ 
and is kept at the same value for a period of $t_2$. 
\begin{figure}
\centerline{\epsfxsize=8cm
\epsffile{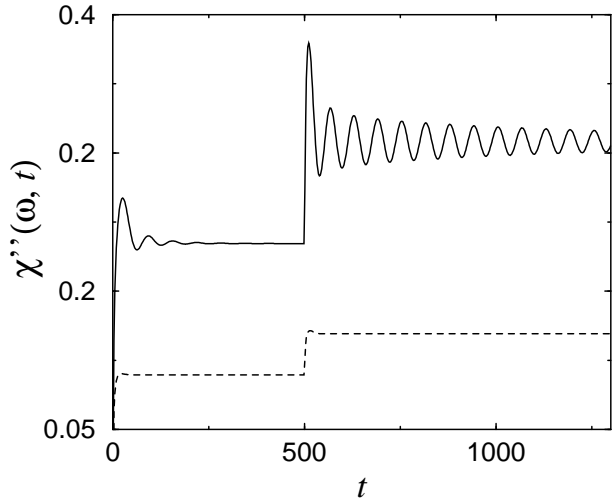}
}
\caption{The out-of-phase susceptibility of the two-state random 
energy model ($T_c = 1$) when the temperature is lowered at $t=500$ from
 $T=0.48$ to $T=0.28$ (the solid line) and from $T=1.0$ to $T=0.8$ (the
 dashed line). The frequency $\omega$ is $0.1$ and the out-of-phase 
susceptibility is averaged over 100 samples. 
The abrupt jump and the subsequent relaxation, i.e. the rejuvenation effect,
 is clearly seen when the temperature is low.}
\label{fig2}
\end{figure}
Fig.~\ref{fig2} shows the result of numerical calculations for 2S-REM when 
$T=0.48 \ \mbox{and} \ T=1.0, \triangle T=0.2, t_1=500$. 
We can clearly see the rejuvenation effect when $T=0.48$, 
i.e. the abrupt jump and the subsequent relaxation of the 
out-of-phase susceptibility when the temperature is changed.
After the jump, $\chi''(\omega,t)$ relaxes to the new steady-state value at 
the temperature $T-\triangle T$ with oscillation. 
As derived in Appendix, 
the steady-state value when the frequency $\omega$ is low compared to
$\tau_0$ behaves as 
\begin{equation}
\lim_{t \rightarrow \infty} \chi''(\omega,t) \simeq \left\{
\begin{array}{ll} \frac{1}{k_B T} \int^{\infty}_0 d\tau [C(\tau)-B] \omega & 
\mbox{when} \ \alpha > 1/2 \\
-\frac{1}{k_B T} 4 A \alpha^2 \Gamma(-2\alpha) \sin(\pi \alpha) 
\omega^{2 \alpha} & \mbox{when} \ \alpha < 1/2, 
\end{array}
\right.
\label{chi_equi}
\end{equation} 
where the definition of the constants $A$ and $B$ are given in eq.~(\ref{c}).
This result implies that the behavior of the steady-state value changes when 
$\alpha = 1/2$.
The qualitatively same result for the 2L-REM is given in Fig.~\ref{fig3}. 
\begin{figure}
\centerline{\epsfxsize=8cm
\epsffile{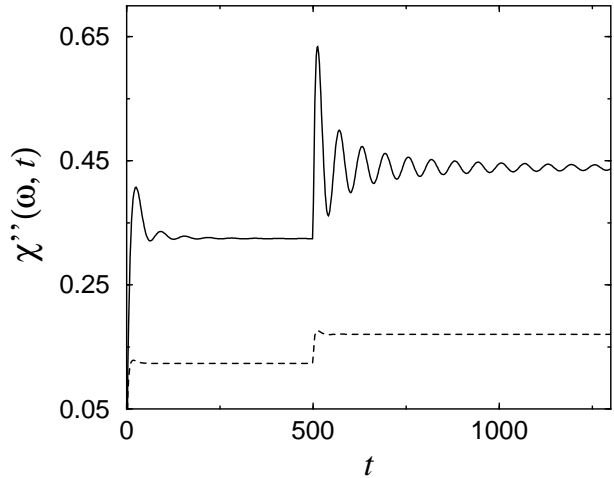}
}
\caption{The out-of-phase susceptibility of the two-layer random 
energy model ($T_c(1) = 0.4, T_c(2)=1.0$) when the temperature is
 lowered at $t=500$ from
 $T=0.48$ to $T=0.28$ (the solid line) and from $T=1.0$ to $T=0.8$ (the
 dashed line). The frequency $\omega$ is $0.1$ and the out-of-phase 
susceptibility is averaged over 100 samples. 
The abrupt jump and the subsequent relaxation, i.e. the rejuvenation effect,
 is clearly seen when the temperature is low.}
\label{fig3}
\end{figure}

\begin{figure}
\centerline{\epsfxsize=8cm
\epsffile{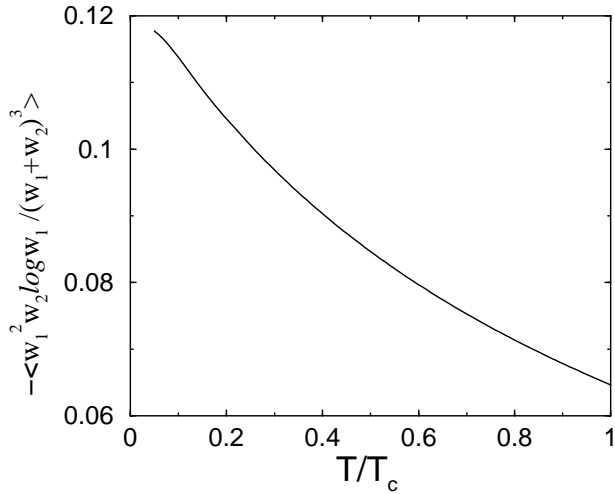}
}
\caption{The $T_c$-dependence of the speed of the change of the out-of-phase 
susceptibility when the temperature is lowered slightly. This plot shows
 that the speed is a increasing function of $T_c/T$ and the abrupt change of
 the out-of-phase susceptibility becomes very steep when the 
state energy deviates broadly or the temperature is low.}
\label{fig4}
\end{figure}

As shown in Fig.~\ref{fig2} and Fig.~\ref{fig3}, 
the rejuvenation effect is clearly seen when the temperature is low.
We show that the abrupt change becomes steep when 
the temperature is low or the disorder of the state energy is strong, 
which is the consequence of sensitivity of equilibrium state to temperature.
Since the qualitative behavior of the out-of-phase susceptibility of the 
2S-REM is same as that of the 2L-REM, we consider the 2S-REM for 
simplicity. 
Since the time lapse of the first stage, $t_1$, is sufficiently long, 
the system is at the steady state.
By solving the master equation when $h$ is small, 
the probability distribution of the steady state at the temperature $T$ is 
derived to the order $O(h)$ as 
\begin{equation}
\left( \begin{array}{c} P_1(t_1) \\ P_2(t_1) \end{array} \right)
\simeq \frac{1}{w_1+w_2} \left( \begin{array}{c} w_2 \\ w_1 
\end{array} \right)+\frac{2w_1 w_2 \exp(-i \omega t_1)}{k_B T(w_1+w_2)
(i \omega-w_1-w_2)} h \left( \begin{array}{c} 1 \\ -1 \end{array} \right).
\end{equation}
From this expression and the master equation eq.~(\ref{master1}), 
the time derivative of the 
out-of-phase susceptibility immediately after the change of the temperature 
is given by 
\begin{equation}
\dot{\chi}''(\omega, t_1^+) \equiv 
\lim_{t \searrow t_1} \frac{d}{dt} \chi''(\omega, t) \simeq 
\frac{4}{k_B T}\frac{w_1 w_2 [w_1+w_2-w_1^{T/(T-\triangle T)}-
w_2^{T/(T-\triangle T)}]}{(w_1+w_2)^3}\omega
\end{equation}
where $w_1$ and $w_2$ are the jump rates before the temperature is changed 
and we assume that the frequency $\omega$ is low.
Assuming that the temperature variation $\triangle T/T$ is small, 
the time derivative of disorder-averaged $\chi''$ is given to the order 
$O(\triangle T/T)$ as 
\begin{equation}
<\dot{\chi}''(\omega, t_1^+)> \simeq -\frac{8 \omega}{k_B T} 
\left<\frac{w_1^2 w_2 \log w_1}{(w_1+w_2)^3}\right> \triangle T/T.
\label{chi}
\end{equation}
The result of numerical calculation of the expression eq.~(\ref{chi}) 
is given in the Fig.~\ref{fig4}. 
We see that the time derivative, $<\dot{\chi}''(\omega, t_1^+)>$, is 
a monotonously increasing function of $T_c/T$. 
Since the deviation of the state energy is a monotonously increasing
function of $T_c$, the above result implies that the abrupt change of
the out-of-phase susceptibility becomes steep when the disorder of the 
state energy is strong or the temperature is low.

\section{DISCUSSION AND CONCLUSIONS}

Summarizing our result, we have considered the out-of-equilibrium 
behavior of the two-state random energy model (2S-REM) 
and its extension with a hierarchy of two levels, i.e. 2L-REM. 
We have shown that these models of energy landscape with very small number 
of local minima can reproduce the out-of-equilibrium phenomena observed in 
glassy materials, especially the rejuvenation effect due to sensitivity of 
equilibrium state to temperature. 
It suggests that some of the interesting phenomena observed in 
glassy materials can be observed in an assembly of small systems with 
relaxation times distributed broadly even if the number of 
local minima is small. 

We close by discussing a future issue to be solved in terms of the 
rugged free-energy landscape picture. 
Recently, it has been found that the phenomenological models based on the 
rugged free-energy landscape picture can explain at least qualitatively 
many of significant dynamical properties of the glassy materials.~
\cite{Bouchaud95,Sasaki00}
However, there is an exception; violation of the fluctuation-dissipation 
theorem described with the violation ratio 
$X(t,t') \equiv k_B T R(t,t')/\partial_{t'}C(t,t')$, where $R(t,t')$ denotes 
the response function, observed in spin-glasses and other glassy materials.~
\cite{Cugliandolo93,Parisi97} 
For example, it is easily shown 
for the models analyzed in \citen{Bouchaud95,Sasaki00}) that the 
violation ratio $X(t,t')$ is given as 
$X(t,t') = t/t'$ from the scaling of the correlation function 
$C(t,t')=\hat{C}(t/t')$ and the relation known to hold for these models 
($k_B T R(t,t') = -\partial_t C(t,t')$)~\cite{Sasaki00}. 
Hence, these models cannot reproduce the constant violation factor 
observed in the p-spin spherical spin-glasses and in the 
numerical studies of the structural glasses~\cite{Parisi97}. 
It would be interesting to create a phenomenological model with 
the rugged free-energy landscape picture be able to reproduce violation of the 
fluctuation-dissipation theorem 
in order to understand the physical meanings of the violation factor 
$X(t,t')$.~\cite{Cugliandolo94}

\section*{Acknowledgements}
I wish to acknowledge valuable discussions with Takashi Odagaki, 
Akira Yoshimori and Jun Matsui.

\appendix
\label{appendix}

\section{derivation of the steady-state value of 
the out-of-phase susceptibility}

Here, we evaluate the steady-state value of the out-of-phase susceptibility. 
We start with the expression of the susceptibility $\chi(\omega,t)$ 
in terms of the response function $R(t,t')$; 
\begin{equation}
\chi(\omega, t) = \int_0^t dt' R(t,t') \exp[i \omega (t-t')].
\label{chi_in_R}
\end{equation}
Noting that the asymptotic form of the correlation function 
is given in eq.~(\ref{c}) 
and the response function is given by the time 
derivative of the correlation function, i.e. $-\partial_{t}C(t,t')/k_B T$, 
when the dynamics for the REM is trap type, i.e. eq.~(\ref{jump_rate}) 
\cite{Sasaki00}, we can evaluate the steady-state value of the 
susceptibility from eq.~(\ref{chi_in_R}).

At first, we divide the right hand side of eq.~(\ref{chi_in_R}) into 
two parts as 
\begin{equation}
\chi(\omega,t) = \int_0^{t-t^{1-b}}dt' R(t,t')e^{i \omega (t-t')}+
\int^t_{t-t^{1-b}}dt' R(t,t')e^{i \omega (t-t')},
\end{equation}
where we assume $0<b<1$.
By using the relation $k_B T R(t,t') = -\partial_t C(t,t')$ and the 
asymptotic form of the correlation function, eq.~(\ref{c}), 
we can evaluate the first term as 
\begin{equation}
|\mbox{the first term}| \leq \frac{2 A \alpha}{k_B T}
[t^{2\alpha(b-1)}-t^{-2\alpha-b}].
\end{equation}
Hence, the first term becomes $0$ in the long time limit, 
$t \rightarrow \infty$.
Next, we evaluate the second term. 
Since $t-t' \ll t'$ in the region of the integration, 
the time translational invariance holds, i.e. $R(t,t')$ is the 
function of the time difference $t-t'$.
Rewriting the response function as $R(t-t')$, 
the second term is given as 
\begin{equation}
\mbox{the second term} = \int_{t-t^{1-b}}^t dt' R(t-t') e^{i \omega (t-t')} 
= \int_0^{t^{1-b}}d\tau R(\tau) e^{i \omega \tau}.
\end{equation}
Hence, the second term becomes in the long time limit 
$
\int_0^{\infty} d\tau R(\tau) e^{i \omega \tau}.
$
Gathering these results, the out-of-phase susceptibility is given 
in the long time limit as 
\begin{equation}
\lim_{t \rightarrow \infty} \chi''(\omega,t) = \int_0^{\infty}d\tau R(\tau) 
\sin(\omega \tau) = -\frac{1}{k_B T} \int_0^{\infty} d\tau 
\frac{dC(\tau)}{d\tau} \sin(\omega \tau).
\end{equation}
Then, noting that the decay of the correlation function $C(\tau)$ is faster 
than $1/\tau$ when $\alpha >1/2$ and is slower than $1/\tau$ when 
$\alpha < 1/2$, by using the asymptotic form of the correlation function 
eq.~(\ref{c}) we reach the expression eq.~(\ref{chi_equi}).


\begin{thebibliography}{99}
\bibitem{Frauenfelder91}H. Frauenfelder, S. G. Sligar and P. G. Wolynes: 
Science {\bf 254} (1991) 1598.
\bibitem{Vincent96}For a review, see E. Vincent, H. Hammann, M. Ocio, J.-P. 
Bouchaud and L. F. Cugliandolo: in {\it Proceedings of 
the Sitges Conference on Glassy systems}, edited by E. Rubi 
(Springer, Berlin, 1996); cond-mat/9607224.
\bibitem{Vincent95}F. Lefloch, J. Hammann, M. Ocio and 
E. Vincent, Europhys. Lett. {\bf 18} (1992) 647; J. O. Andersson, 
J. Mattsson and P. Nordblad, Phys. Rev. B {\bf 48} (1993) 1397; 
E. Vincent, J.-P. Bouchaud, J. Hammann and F. Lefloch, 
Phil. Mag. B {\bf 71} (1995) 489. 
\bibitem{Jonason98}K. Jonason, E. Vincent, J. Hammann, J.-P. Bouchaud and 
P. Nordblad, Phys. Rev. Lett. {\bf 81}, (1998) 3243.
\bibitem{Bouchaud95}J.-P. Bouchaud and D. S. Dean, J. Phys. I France {\bf 5} 
(1995) 265; M. Sasaki and K. Nemoto, J. Phys. Soc. Jpn. {\bf 69} (2000) 2283
; cond-mat/0010443.
\bibitem{Sasaki00}M. Sasaki and K. Nemoto, J. Phys. Soc. Jpn. 
{\bf 69} (2000) 3045.
\bibitem{Cugliandolo93}L. F. Cugliandolo and J. Kurchan, Phys. Rev. Lett. 
{\bf 71} (1993) 173; J. Phys. A {\bf 27} (1994) 5749; 
L. F. Cugliandolo, J. Kurchan and L. Peliti, Phys. Rev. E {\bf 55}
(1997) 3898. 
\bibitem{Parisi97}G. Parisi, Phys. Rev. Lett. {\bf 79} (1997) 3660; 
J. -L. Barrat and W. Kob, Europhys. Lett. {\bf 46} (1999) 637.
\bibitem{Cugliandolo94}There are some surveys in this direction. 
For example, 
L. F. Cugliandolo, J. Kurchan and G. Parisi, J. Phys. I {\bf 4} (1994) 1641; 
L. Laloux and P. L. Doussal, Phys. Rev. E {\bf 57} (1998) 6296.
\end{thebibliography}
\end{document}